\documentclass[%
 twocolumn,
 prl,
superscriptaddress,
showkeys,
 amsmath,amssymb,
 aps,
]{revtex4-2}

\usepackage{mathtools}
\usepackage{textcomp}
\usepackage{mathrsfs}

\usepackage{amsfonts}
\usepackage{hyperref}
\usepackage{nicefrac}
\usepackage{graphicx}% Include figure files
\usepackage{dcolumn}% Align table columns on decimal point
\usepackage{bm}% bold math

\begin{document}

\preprint{APS/123-QED}

\title{Metabolic activity controls the emergence of coherent flows in microbial suspensions}% Force line breaks with \\

\author{Alexandros A. Fragkopoulos}
\affiliation{Experimental Physics V, University of Bayreuth, Universit\"atsstr.\ 30, D-95447 Bayreuth, Germany}
\affiliation{Max Planck Institute for Dynamics and Self-Organization (MPIDS), Am Fa{\ss}berg 17, D-37077 G\"{o}ttingen, Germany}

\author{Florian B\"{o}hme}
\affiliation{Experimental Physics V, University of Bayreuth, Universit\"atsstr.\ 30, D-95447 Bayreuth, Germany}

\author{Nicole Drewes}
\affiliation{Max Planck Institute for Dynamics and Self-Organization (MPIDS), Am Fa{\ss}berg 17, D-37077 G\"{o}ttingen, Germany}

\author{Oliver B\"{a}umchen}
\email[]{oliver.baeumchen@uni-bayreuth.de}
\affiliation{Experimental Physics V, University of Bayreuth, Universit\"atsstr.\ 30, D-95447 Bayreuth, Germany}
\affiliation{Max Planck Institute for Dynamics and Self-Organization (MPIDS), Am Fa{\ss}berg 17, D-37077 G\"{o}ttingen, Germany}

\date{\today}% It is always \today, today,

\begin{abstract}
Photosynthetic microbes have evolved and successfully adapted to the ever-changing environmental conditions in complex microhabitats throughout almost all ecosystems on Earth.
In the absence of light, they can sustain their biological functionalities through aerobic respiration, and even in anoxic conditions through anaerobic metabolic activity.
For a suspension of photosynthetic microbes in an anaerobic environment, individual cellular motility is directly controlled by its photosynthetic activity, i.e.\ the intensity of the incident light absorbed by chlorophyll.
The effects of the metabolic activity on the collective motility on the population level, however, remain elusive so far.
Here, we demonstrate that at high light intensities, a suspension of photosynthetically active microbes exhibits a stable reverse sedimentation profile of the cell density due to the microbes' natural bias to move against gravity.
With decreasing photosynthetic activity, and therefore suppressed individual motility, the living suspension becomes unstable giving rise to coherent bioconvective flows.
The collective motility is fully reversible and manifests as regular, three-dimensional plume structures, in which flow rates and cell distributions are directly controlled via the light intensity.
The coherent flows emerge in the highly unfavourable condition of lacking both light and oxygen and, thus, might help the microbial collective to expand the exploration of their natural habitat in search for better survival conditions.
\end{abstract}

\maketitle

%\tableofcontents

\section*{Introduction}

Photoautrophic microorganisms colonize almost all natural habitats and ecosystems on our planet, by adapting to their frequent and sometimes harsh environmental changes \cite{raven2003adaptation}. 
Specifically, spatio-temporal alterations of the light intensity \cite{singh1996reactivation,xue1996interactions} and oxygen (O$_2$) concentration \cite{atteia2013anaerobic,yang2015algae} exert profound effects on the metabolic pathways employed by the cells to produce life-sustaining chemical energy.
Even though there exists a diverse set of metabolic pathways, these can generally be grouped into three main categories: photosynthesis, aerobic respiration, and anaerobic fermentation \cite{grossman2023chlamydomonas}.
The latter allows for the cells to thrive in a light- and oxygen-depleted environment, termed dark anoxia, \cite{mus2007anaerobic}, which can occur naturally in habitats providing little gas permeability such as soil \cite{garcia2003small}.
These conditions can also occur during water eutrophication, where the rapid growth of microorganisms results in the depletion of oxygen in the dissolved water \cite{anderson2002harmful,glibert2017eutrophication}.

Understanding and controlling the implications of metabolic pathway switches is not only pivotal for ecological reasons, but also for technological applications.
Microalgae, and specifically the model organism \textit{Chlamydomonas reinhardtii}, represent microbial systems to produce clean and renewable hydrogen (H$_2$) \cite{esquivel2011efficient,torzillo2015advances}.
While a minor production of H$_2$ is observed during fermentation \cite{gfeller1984fermentative,mus2007anaerobic}, the highest yield can be achieved through hydrogenase during photosynthesis \cite{forestier2003expression,esquivel2011efficient,torzillo2015advances}.
Hydrogenase in \textit{Chlamydomonas} is highly sensitive to O$_2$, which is a product of photosynthesis, and thus H$_2$ is naturally inhibited during photosynthesis.
As a result, a combination of anoxic conditions with alterations between photosynthesis and darkness seems to attain optimal H$_2$ production \cite{esquivel2011efficient,williams2014mechanistic,torzillo2015advances}.

\begin{figure*}%[tbhp]
\centering
\includegraphics{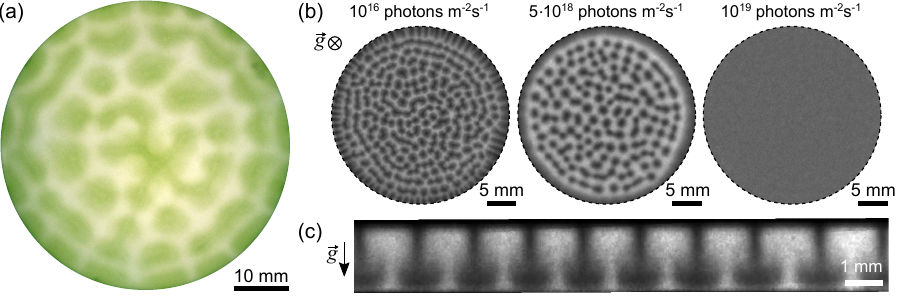}
\caption{
\textbf{Light-controlled emergence of coherent flows in living \textit{C.~reinhardtii} suspensions.} 
(\textbf{a}) Top-view of a suspension of \textit{C.~reinhardtii} under aerobic conditions and homogeneous white light in a Petri dish with cell concentration $\rho_0=8\cdot10^7$\,cells mL$^{-1}$ and height $3$\,mm.
(\textbf{b}) Top-view: A suspension of \textit{C.~reinhardtii} under anaerobic conditions and red light at different light intensities, confined in an air-tight 3D compartment (height $1$\,mm, cell concentration $\rho_0=8\cdot10^7$\,cells mL$^{-1}$).
The suspension displays a homogeneous cell density at high light intensities (right), while convective patterns are observed at reduced light intensities (left and middle).
Effects are completely reversible when switching between different light intensities $I$.
(\textbf{c}) Side-view: Structure of the convective rolls of a suspension of \textit{C.~reinhardtii} under anaerobic conditions confined in an air-tight quasi-2D compartment ($\rho_0=8\cdot10^7$\,cells mL$^{-1}$,  $I=4.6\cdot10^{18}$\, photons m$^{-2}$ s$^{-1}$).
See also Movies~S1 and S2 for the onset of bioconvection in both types of compartments.
}
\label{fig:fig1}
\end{figure*}

Metabolic pathway analysis has been used in the study of physiological capacities and features of metabolic networks \cite{cakir2006integration,kromer2006metabolic,unrean2011metabolic}.
The relation, however, between the metabolic pathways and the cell motility, and by extension the emergence and appearance of collective behavior of a living suspension, remains largely unexplored \cite{fragkopoulos2021self}.
Self-organized coherent structures in dense microbial suspensions that arises due to the natural tendency of microorganisms to swim against gravity are often summarized under the term `bioconvection', which is known for depending on the single-cell motility \cite{hill2005bioconvection,sengupta2017phytoplankton,bees2020advances}.

Here, we show that in anaerobic conditions, the spatio-temporal characteristics of coherent flows change and may even reversibly emerge or disappear, depending on the photosynthetic activity of the cells.
As a result of the anaerobic conditions, the microbial motility is directly controlled by the light intensity for chlorophyll absorption. 
Phototactic effects, i.e.\ a biased collective motility in light gradients \cite{dervaux2017light}, can be safely ruled out in this scenario through the use of red light and/or photoreceptor (channelrhodopsins) knockout cells \cite{sineshchekov2002two}.
The spatio-temporal morphology, cell distributions and cell circulation rates within the coherent flow structures are regulated by the light intensity, demonstrating a direct link between the activity of metabolic pathways and single-cell as well as collective microbial motility.

\section*{Light controlled coherent flows in suspensions of photosynthetic microbes}

Many species within the diverse group of photosynthetic microorganisms perform negative gravitaxis, i.e.\ exhibit the tendency to move against the gravitational field \cite{roberts2006mechanisms}.
This form of taxes is essential for their survival, since it allows the cells to increase their chances of locating regions in their environment with more optimal light and aeration conditions.
Negative gravitaxis is believed to emerge from the combination of the bottom-heaviness and the shape asymmetry of the cells \cite{roberts2006mechanisms}.
The term `bottom-heaviness' refers to the inhomogeneous density distribution within a cell, with the chloroplast at the rear of the cell being denser, causing a gravitational torque that aligns the cells against gravity \cite{kage2020shape}.
Similarly, the `shape asymmetry' refers to the asymmetric drag force due to the presence of the flagella at the anterior of the cells, causing a similar torque during cell sedimentation \cite{kage2020shape,roberts2002gravitaxis}.
Due to negative gravitaxis, the cells tend to form a dense layer at the top surface, with the vertical cell density to theoretically follow an exponential decrease toward deeper regions \cite{hill1989growth}.
This inverse sedimentation profile is only stable for sufficiently shallow containers, and may become unstable above a critical depth.
The latter results in the emergence of distinct bioconvective density patterns \cite{hill1989growth,williams2011tale}, as seen in macroscopic experiments, performed in an open Petri dish, shown in Fig.~\ref{fig:fig1}a.
These bioconvective patterns are stable and display a typical length scale of over 1\,mm.

We employ custom-made setups to study suspensions of motile \textit{C.~reinhardtii} cells confined in two types of air-tight compartments, see Fig.~S1a in the SI.
First, cylindrical compartments with a diameter of $30$\,mm and height ranging between $0.5-1.0$\,mm are used to limit edge effects and study the pattern formation in top view, see Fig.~S1b in the SI.
Second, a rectangular cuboid with $30$\,mm length and $2$\,mm width and height is used to characterize the coherent flows in side view, see Fig.~S1c in the SI.
The width is sufficiently small to allow the development and obsevation of single plumes along the width of the compartment.
These two compartments will be from now referred to as ``3D'' and ``quasi-2D'' compartments.
All experiments are performed using red light (wavelength $\lambda = 660\pm20$\,nm) in order to safely exclude phototactic effects \cite{berthold2008channelrhodopsin} and surface attachment of the cells \cite{kreis2018adhesion}. 
For consistency the (global) cell density $\rho_0$ is kept constant for all experiments at  about $8\cdot10^7$\,cells mL$^{-1}$. Additionally, we safely ruled out effects of phototaxis by testing double-knockout mutants of channelrhodopsin-1 and -2, see Fig.~S2 in the SI.

When the dense suspension of \textit{C.~reinhardtii} cells is exposed to a light intensity of $I=10^{19}$\, photons m$^{-2}$ s$^{-1}$, we find that the cells appear homogeneously distributed when viewed from the top, see Fig.~\ref{fig:fig1}b(right). 
By gradually reducing the light intensity, and thus diminishing the photosynthetic activity, the cells are subjected to ``self-anaerobization'' \cite{fragkopoulos2021self}.
During this process the cells consume the dissolved O$_2$ and exhibit short- and long-term changes on the transcription of metabolic enzymes \cite{hemschemeier2013copper}, while the boundary conditions inhibit replenishment of the consumed O$_2$.
We find that under these anaerobic conditions, the same suspension of motile cells forms bioconvection plumes, starting from the edges of the compartment and moving towards the center, until the whole compartment is filled, see Fig.~\ref{fig:fig1}b(left,middle) and Movie~S1.
By using light intensity as a control parameter, we can trigger the formation and disappearance of large-scale coherent flows, and also alter their spatio-temporal morphology.
All these effects are completely reversible when reverting the light conditions for the same population of microorganisms.

Even though the microbial suspension at high light intensity appears homogeneous from the top, as discussed, the vertical distribution of cells is not homogeneous due to negative gravitaxis, c.f.\ Fig.~\ref{fig:fig3}b.
To observe modulations of the density along the axis of gravity, we use the side-view imaging offered by the quasi-2D experiments.
This setup allows us to quantify the quasi-stationary sedimentation profile at the ``homogeneous'' (high light intensity) state, and also the spatio-temporal evolution as well as the steady-state of the cell density distributions and flow rates within a plume (low light intensities), as depicted in Fig.~\ref{fig:fig1}c and Movie~S2.

\section*{Morphological characterization of emergent light-controlled coherent flows}

In order to systematically characterize the emergence and spatio-temporal characteristics of large-scale flow patterns, we perform experiments by first exposing the cell suspension to high light intensity for $10$\,minutes to homogenize the density and achieve a quasi-steady state.
Afterwards, the light intensity is reduced to the desired lower light intensity, and the spatio-temporal evolution of the system is recorded for $15-20$\,minutes.
We then employ a Fourier transform analysis to characterize the evolution of the flow patterns, see Methods for further details.

\begin{figure}%[tbhp]
\centering
\includegraphics{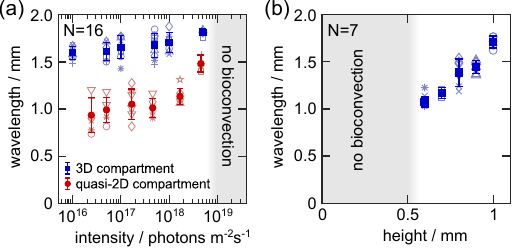}
\caption{\textbf{Light intensity and compartment height control the emergence and appearance of coherent microbial flows.} 
(\textbf{a}) The wavelength of the pattern as a function of the incident light intensity for suspensions with cell concentration $\rho_0=8\cdot10^7$\,cells mL$^{-1}$ for 3D compartments (blue squares) and quasi-2D compartments (red circles).
The height of the compartment is $1$\,mm. 
(\textbf{b}) The wavelength of the pattern for 3D compartments as a function of the height of the compartment for a suspension with cell concentration $\rho_0=8\cdot10^7$\,cells mL$^{-1}$ and light intensity $I=10^{16}$\, photons m$^{-2}$ s$^{-1}$.
Open symbols indicate technical replicates comprising 1 to 8 repetitions at each light intensity.
Filled symbols and their error bar indicate the mean and standard deviation of the $N$ independent biological replicates.
Experiments with different light intensities were performed in randomized order.
}
\label{fig:fig2}
\end{figure}

The peak in the power spectrum $S(q,t)$ corresponds to the most prominent distance within the pattern.
By extracting the wavenumber, $q_{max}$, that corresponds to this peak, a wavelength can be assigned to the pattern as $\lambda=q_\mathrm{max}^{-1}$.
This wavelength was measured at different light intensities for both 3D and quasi-2D compartments, see Fig.~\ref{fig:fig2}a.
In both cases, we do not observe any bioconvection for light intensities $I\ge10^{19}$\, photons m$^{-2}$ s$^{-1}$, while bioconvection always occurs for $I\le 2\cdot10^{18}$\, photons m$^{-2}$ s$^{-1}$.
Experiments around $I=5\cdot10^{18}$\, photons m$^{-2}$ s$^{-1}$ exhibit bistability, with experimental repetitions using the same culture arbitrarily exhibiting bioconvection or the homogeneous state.
For 3D compartments, a systematic measure of the wavelengths shows that there is at most a weak dependence of the wavelength on the incident light intensity, see Fig.~\ref{fig:fig2}a.
However, there is a substantial discrepancy on the values of the wavelengths between the two geometries.
For 3D compartments with a height of $1$\,mm, the wavelength of the bioconvection pattern is $\lambda=1.7\pm0.1$\,mm.
In the case of the quasi-2D compartment with double the height at $2$\,mm, the wavelength is measured to be significantly smaller at $\lambda=1.1\pm0.1$\,mm.
Consequently, the confinement of the suspension along one dimension has a profound effect on the fastest unstable mode, such that it is significantly reducing the wavelength of the instability.
The wavelength also exhibits a weak dependence on the incident light, except close to the critical light intensity, at which we observe a significant increase.

\begin{figure*}%[tbhp]
\centering
\includegraphics{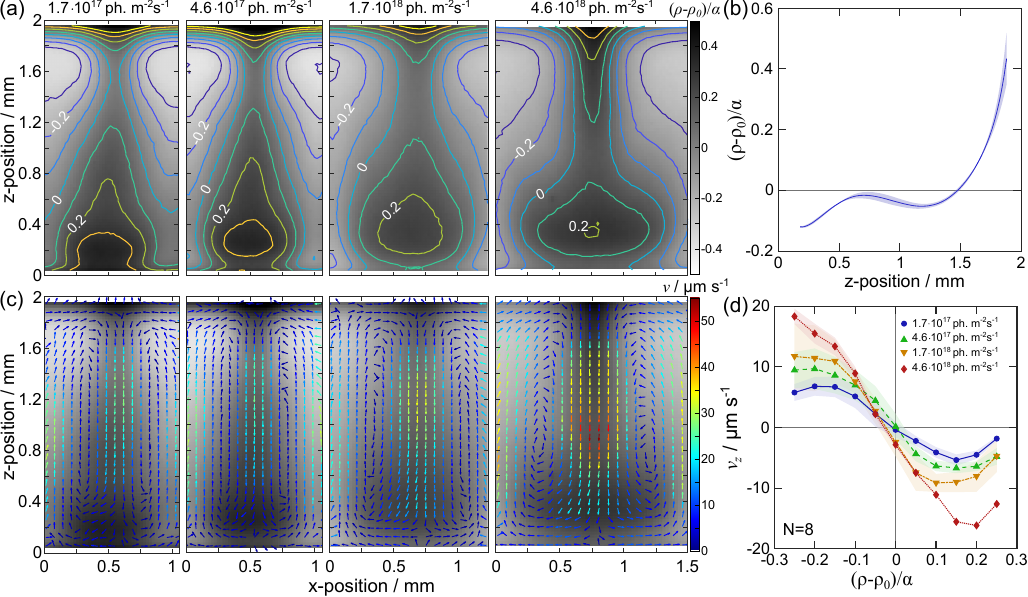}
\caption{
\textbf{Density distribution and cell flow organization within plumes in \textit{C.~reinhardtii} suspensions.} 
(\textbf{a}) Vertical cross-sections of the relative cell density $(\rho-\rho_0)/\alpha$ of stable bioconvective plumes at different light intensities.
%Experiments shown are from different biological replicates.
Contour lines represent the isolines for relative cell densities $(\rho-\rho_0)/\alpha$.
(\textbf{b}) Inverse sedimentation profile: The relative cell density $(\rho-\rho_0)/\alpha$ at high light intensity ($I=4.9\cdot10^{19}$\, photons m$^{-2}$ s$^{-1}$) is plotted as a function of the $z$-coordinate with respect to the bottom of the container.
At this light intensity, the suspension exhibits an inverted sedimentation profile and no coherent flows.
(\textbf{c}) Flow velocity fields within stable bioconvective plumes are measured using Particle Image Velocimetry (PIV) for the experiments and light intensities displayed in panel (a).
(\textbf{d}) Velocity-density coupling: The vertical component of the cell flow, $v_z$, is displayed as a function of the relative cell density $(\rho-\rho_0)/\alpha$ for the flow velocity fields displayed in panel (c).
The value of $v_z$ is given as the average vertical velocity for the same density $(\rho-\rho_0)/\alpha$ with a bin size of 0.5, with a positive $v_z$ indicating an upward cell flow.
All data are for cell suspensions in a quasi-2D compartment with an average cell density $\rho_0=8\cdot10^7$\,cells mL$^{-1}$.
Filled symbols and their error indicate the mean and standard deviation of $N$ independent biological replicates.
Experiments with different light intensities were performed in randomized order.
}
\label{fig:fig3}
\end{figure*}

This confinement effect is further emphasized by analyzing the dependence of the wavelength with the height of the compartment.
Figure~\ref{fig:fig2}b shows a monotonic increase of the wavelength of the instability with the compartment height for 3D geometries.
This observation is in accordance with previous theoretical work on the subject using no-slip boundary conditions for both surfaces \cite{pedley1988growth}.
As in previous work, we also observe a critical height below which no convective flows are observed \cite{hill2005bioconvection,williams2011tale}.
Specifically, in our case bioconvection never forms for $h\le0.5$\,mm.
Furthermore, experiments at $h=0.6$\,mm are metastable, where bioconvection appears after lowering the light intensity, but always disappears within the course of the experiment.

Now, we use the power spectrum to quantify the temporal dynamics of the pattern formation.
More specifically, the temporal evolution of the maximum of power spectrum, $S_\mathrm{max}(t)$, is used to measure the dynamic properties of the instability.
Since the experiments start in the homogeneous state, $S_\mathrm{max}(t)$ has initially a small value. 
As the instability develops, an initial regime can be identified in which $S_\mathrm{max}(t)$ increases exponentially, before it deviates due to non-linear contributions, and eventually reaches a plateau.
An example of $S_\mathrm{max}(t)$ in a 3D compartment is shown in Fig.~S3c in the SI.
From these measurements, we identify the onset time, $t_\mathrm{on}$, which represents the time required for bioconvection to start after the light intensity is lowered.
We find that the onset time is $t_\mathrm{on}=220\pm60$\,s (from $N=14$ biological replicates with 64 overall repetitions) and independent of the light intensity; see Fig.~S4a in the SI for further details.
The suspension needs to self-anaerobize, i.e.\ consume all the O$_2$ within the medium, before bioconvection may be initiated and since we use the same density in the experiments, we expect $t_\mathrm{on}$ to be constant for our experiments.
Finally, we fit the exponential increase after the onset of the instability as $S_\mathrm{max}(t)\propto e^{\omega\,t}$, to extract the growth rate $\omega$, as shown in Fig.~S3c in the SI.
We find that the growth rate $\omega=(4\pm2)\cdot 10^{-2}$\,s$^{-1}$ (from $N=14$ biological replicates with 64 overall repetitions) is independent of the light intensity and also of type of compartment used; see Fig.~S4b for further details.
Taking these results together we conclude that confinement and light intensity control the appearance, but not the time scales of the evolution of the instability.

\section*{Density distributions and flow fields within coherent flows}

We now systematically characterize the effect of light intensity on the cell density distributions and the cell flow organization within plumes in the quasi-steady state.
By applying eq.~\ref{eq:LB3} derived in the Methods, we estimate the relative cell density within stable bioconvective plumes at different light intensities based on the quasi-2D experiments.
We observe that the cell distribution changes significantly at different light conditions, with the center of the plume (region of highest cell density) progressively moving upward with increasing light intensity, as depicted in Fig.~\ref{fig:fig3}a.
We also observe that the range of relative cell densities becomes narrower as the light intensity increases. 
Specifically, the highest observed relative cell density at $I= 1.7\cdot10^{17}$\, photons m$^{-2}$ s$^{-1}$ is $(\rho-\rho_0)/\alpha\approx0.4$, while at $I= 4.6\cdot10^{18}$\, photons m$^{-2}$ s$^{-1}$ is $(\rho-\rho_0)/\alpha\approx0.2$.
At sufficiently high light intensity the bioconvection disappears, and an inverse sedimentation profile is instead acquired, as shown in Fig.~\ref{fig:fig3}b.

Complementary to the experimental quantification of cell density distributions, we also established particle image velocimetry (PIV) methods to measure the flow of cells within the plumes \cite{adrian2011particle}.
Conventional PIV entails that a fluid is seeded with tracer particles and consecutive images of the flow with the particles are correlated in order to measure the local fluid flow.
%In the case of fluid visualization, the particles are required to follow the fluid flow in order to adequately measure it.
Here, we use the cells as the tracer particles, which due to the self-propelled nature of the cells do not necessarily follow the flow of the surrounding fluid.
Thus, we rather measure the cell flow, which is the combination of the fluid flow coupled with the average motility of the cells with respect to the surrounding fluid.

For all light intensities, we observe that the cells accumulate in regions of downwelling flow, while the relative density is substantially reduced in regions of upwelling flow as compared to the global cell density, see Fig.~\ref{fig:fig3}c.
This is expected due to gyrotactic effects.
The combination of the gravitactic torque, which aligns the cells against gravity, and the viscous torque, which is the result of the non-zero vorticity field, cause the cells to orient on average towards downwelling flows, a phenomenon known as gyrotaxis \cite{kessler1985hydrodynamic}.
For different light intensities, the circulation rates increases with increasing light intensity, with the maximum cell flow velocity rising from $25$\, \textmu m\,s$^{-1}$ to $55$\, \textmu m\,s$^{-1}$.
These findings are in line with a higher motility of the microbes on the single-cell level at higher light intensities under anaerobic conditions, which will discussed in more detail in the following paragraph.

To further quantify these effects on the population scale, we measure the average vertical cell flow, $v_z$, in regions of comparable cell density as a function of the local cell density, as shown in Fig.~\ref{fig:fig3}d.
We observe the gyrotactic accumulation of cells in regions of downwelling flow, which manifests as the negative correlation between the vertical component of the cell flow and the cell density.
Finally, a faster turnover of the cell flow is observed at high light intensities, with the maximum circulation velocity increasing from $8$\,\textmu m\, s$^{-1}$ to $19$\,\textmu m\, s$^{-1}$ with increasing light intensity.

\begin{figure}%[tbhp]
\centering
\includegraphics{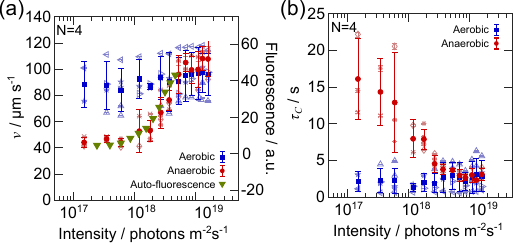}
\caption{
\textbf{Effect of the photosynthetic activity on the single-cell motility.} 
(\textbf{a}) The root-mean-square velocity, $v$, and (\textbf{b}) the velocity correlation time, $\tau_c$, of the cells as a function of light intensity under aerobic (blue circles) and anaerobic (red squares) conditions.
Multiple experiment were performed at different cell densities, ranging between $500-1300$\,cells\,mm$^{-2}$.
(\textbf{a}) Chlorophyll auto-fluorescence (green triangles) as a function of the light intensity represents a proxy of the photosynthetic activity of the cells and correlates well with the cell velocity in anaerobic conditions (red squares). 
See Methods for further details.
Experiments were performed in compartments with a height of $h=25$\,\textmu m in order to confine the motility of the cell to 2D.
For each biological replicate, the trajectories of 1500--4500 cells were recorded and were used to calculate $v$ and $\tau_c$.
Open symbols indicate technical replicates comprising 1 to 3 repetitions at each light intensity.
Filled symbols and their error bar indicate the mean and standard deviation of the $N$ independent biological replicates.
Experiments with different light intensities were performed in randomized order.
}
\label{fig:fig4}
\end{figure}

\section*{Effect of metabolic activity on the single-cell motility}

Linking the emergence and appearance of coherent flows to the activity of the individual constituents requires a quantitative understanding of the cell motility and its connection to the microbial metabolism.
The motion of \textit{C.~reinhardtii} can be modeled as a run-and-tumble (RT) motility \cite{polin2009chlamydomonas}, where the cells exhibit short-time ballistic motion (run) that is interrupted by sudden reorientations (tumble).
The motility can be quantified using the velocity autocorrelation function \cite{caprini2020time}, $C_v(t) = \langle \textbf{v}_i(t=0)\cdot\textbf{v}_i(t)\rangle$, where the average is taken over all tracked cells with $\textbf{v}_i$ representing the velocity of the individual cell $i$. 
These experiments are performed in compartments with a height of $25$\,\textmu m following the protocol outlined in \cite{fragkopoulos2021self}.
In order to dissect the effect of the microbial metabolism, experiments are performed under aerobic (air-permeable compartments) and anaerobic (airtight compartments) conditions, while monitoring the photosynthetic activity using the chlorophyll autofluorescence signal as a proxy.

The velocity autocorrelation $C_v$ follows an exponential decay in all cases (see Fig.~S5 in the SI) and can be expressed as $C_v=v^2e^{-\nicefrac{t}{\tau_\mathrm{c}}}$, where $v=\sqrt{\langle \textbf{v}^2 \rangle}$ is the root-mean-square velocity, and $\tau_c$ denotes the velocity correlation time \cite{fragkopoulos2021self}.
In aerobic conditions, the motility is independent of the light intensity with $v=92\pm16$\,\textmu m\, s$^{-1}$ and $\tau_\mathrm{c} = 2.4\pm1.7$\, s, as shown in Fig.~\ref{fig:fig4}(a,b).
In contrast, the motility of the cells directly depends on the light conditions in the case of anaerobic conditions, i.e.\ the root-mean-squared velocity $v$ monotonically decreases while the velocity correlation time $\tau_\mathrm{c}$ increases with decreasing light intensity, respectively.
These values eventually reach a plateau at $v=46\pm3$\,\textmu m\,s$^{-1}$ and $\tau_\mathrm{c}=15\pm5$\,s for $I< 7\cdot10^{17}$\, photons m$^{-2}$ s$^{-1}$.
For light intensities $I\ge 7\cdot10^{18}$\, photons m$^{-2}$ s$^{-1}$, the motility of the cells is consistent with its values obtained for aerobic conditions.
At the same time we observe that the chlorophyll auto-fluorescence increases with the incident light intensity in the range of intensities where the cell motility changes, as depicted in Fig.~\ref{fig:fig4}a.
Furthermore, the chlorophyll auto-fluorescence plateaus to a minimum for $I< 7\cdot10^{17}$\, photons m$^{-2}$ s$^{-1}$, which is in line with the cell motility.

In summary, single-cell motility assays together with measurements of the chlorophyll autofluorescence of the microbes demonstrate that under anaerobic conditions, the photosynthetic activity directly controls the motility of the cells and thus appears to be responsible for the emergence of coherent flows in such light-regulated living fluids.

\section*{Discussion}
To understand how rather subtle changes of individual microbial motility may induce or suppress population-scale convective instabilities in living fluids, we briefly revisit the state-of-the-art theoretical background for bioconvection.
Continuum theories of gyrotactic organisms characterize such systems using four dimensionless numbers:
\begin{align}
    \textrm{bioconvective Rayleigh number:\ } & R = \frac{g\,\phi\,\Delta\rho\,h^3}{\rho\,\nu\,D}, \label{eq:R}\\
    \textrm{scaled cell speed:\ } & w = \frac{v~h}{D}, \label{eq:w}\\
    \textrm{gyrotactic number:\ } & G = \frac{B\,D}{h^2}, \label{eq:G}\\
    \textrm{Schmidt number:\ } & Sc=\frac{\nu}{D}, \label{eq:Sc}
\end{align}
where $g$ is the gravitational acceleration, $\phi$ is the volume fraction of the suspension, $\Delta \rho=\rho_\mathrm{c}-\rho$ with $\rho_\mathrm{c}$ and $\rho$ the mass density of the cell and the fluid respectively, $\nu$ the kinematic viscosity, $D$ the diffusion constant of the cells, and $B$ the time for a cell to orient with gravity due to gravitaxis \cite{hill1989growth}.
Similar to the classical Rayleigh-B\'{e}nard instability in a nonliving fluid, the suspension is unstable and thus is expected to be prone to bioconvective instabilities above a critical Rayleigh number, i.e.\ for $R>R_\mathrm{c}(w,G,Sc)$.

In the following, we derive the relevant dimensionless numbers specifically for our system of photoactive microbes. 
We retrieve the diffusion constant $D$ of the cells by using the velocity autocorrelation functions as
\begin{equation}
    D = \frac{1}{d}\int_0^\infty C_v(t)\,\mathrm{d}t,
\end{equation}
where $d$ is the dimensionality of the system \cite{fodor2018statistical}.
Since the experiments where performed in a quasi-2D environment, the diffusion constant is given as $D = v^2\tau_c/2$, provided $C_v$ follows the exponential form.
Even though both $v$ and $\tau_c$ change with light under anaerobic conditions, see Fig.~\ref{fig:fig4}, we find that $D$ remains largely independent of the light intensity with a value of $D = (1.6\pm0.5)\cdot 10^4$\,\textmu m$^2$\,s$^{-1}$.
As a result $R$, $G$ and $Sc$ are constant for the different light intensities, with only the scaled velocity $w$ increasing substantially from $5.8$ to $11.5$ with increasing light intensity.
We find the Rayleigh number $R = 640$ for the 3D experiments using the corresponding values of $g=9.8$\,m\,s$^{-2}$, $\phi=2.1\,\%$, $\Delta \rho/\rho = 0.05$ \cite{harris2023}, $h=1$\,mm, $\nu=10^{-6}$ m$^2$\,s$^{-1}$, and $D = 1.6\cdot 10^4$\,\textmu m$^2$\,s$^{-1}$.

Since for this type of living fluid $R$ is a constant, the only way for the suspension to reversibly switch between stable and unstable conditions is a change of the critical Rayleigh number, $R_\mathrm{c}$, between light intensities.
The system exhibits constant $G$ and $Sc$, and thus $R_\mathrm{c}$ may only change if $w$ changes.
For suspensions exhibiting a finite depth with $w>1$, no-slip boundary conditions and an initial inverted sedimentation profile, as in our case, $R_\mathrm{c}\propto w^4$ \cite{hill1989growth}.
Consequently, the value of $R_\mathrm{c}$ is reduced by a factor of 15 between the high and low light intensities in our photoactive living fluid.
As the light intensity and, thus, $w$ decrease, $R_c$ decreases to the point that the Rayleigh number of the system becomes larger than its critical value, i.e.\ $R>R_c$.
As a result the suspension becomes unstable initiating the global formation of bioconvective flows.
The fastest unstable mode, which determines the wavelength of the pattern, is almost independent of $w$ \cite{hill1989growth}, in agreement with our observations.
Unfortunately, we cannot explicitly calculate the actual values of $R_\mathrm{c}$ and the wavelength at different light intensities, since $B$ remains unknown.
It is important to note that this theoretical approach has some limitations, since it ignores reorientations of the cells due to tumbles and the effect of cell-cell interactions.
Such cell-cell interactions are known to cause changes of the microbial motility characteristics, causing the velocity and diffusion constant to be density dependent \cite{fragkopoulos2021self}.

This mechanism is further supported by simulations of bottom-heavy active Brownian particles including steric as well as hydrodynamic interactions, where the active suspension is found to transition from an inverted sedimentation to plumes and finally to regular sedimentation with decreasing particle velocity \cite{ruhle2020emergent}.
However it needs to be noted that in this numerical work \cite{ruhle2020emergent}, the relative size of cells and compartment height differ from our experiments, and the plumes observed are not stable but rather transient states.

In contrast to our experiments, previous work has exclusively focused on aerated suspensions with an air-liquid interface. 
In that case, variation of the illumination intensity of red light does neither alter the formation nor the morphology of the bioconvection patterns for \textit{Chlamydomonas augustae}, a similar photosynthetic microorganisms \cite{williams2011tale}.
For \textit{Chlamydomonas reinhardtii} a reduction of red-light intensity using intermittent short-time exposure to light pulses has been reported to not show an immediate change of the bioconvective pattern, even though such periodic illumination is found to alter long-term pattern rearrangements \cite{kage2013drastic}. 
Here, we use continuous illumination in anaerobic conditions and show that the emergence and disappearance of bioconvective patters are reversibly switchable by light. 

\section*{Conclusion}
We reveal the effect of the metabolism on the individual as well as the collective microbial motility.
Under aerobic conditions, the motility of photosynthetically active microbes remains unaffected by the intensity of the incident light, since cells access oxygen enabling them to perform aerobic respiration.
In contrast, the cell motility exhibits drastic changes with decreasing light intensity under anaerobic conditions, providing a unique control parameter for the microbial motility in the complete absence of phototaxis.
The changes of the metabolic activity and the motility of individual cells exert profound effects on the population dynamics.
Due to the natural tendency of the cells to move against gravity, the living suspension exhibits a stable inverse sedimentation profile at high light intensities, but develops coherent flows as the light intensity decreases sufficiently for the single-cell motility to be altered.
In this system, light provides direct experimental access to the impact of microbial motility on bioconvective flows.
In line with previous analytical as well as numerical work, a decrease of the cell velocity induces the manifestation of bioconvective patterns.
The physical mechanism behind this phenomenon bases on the microbes' preference to accumulate at the top interface at high light intensity and propulsion velocity due to their dominant bottom heaviness and thus upward orientation during swimming.
On the population scale, this effect causes an inverted sedimentation profile within the living suspension.
As the propulsion velocity decreases substantially with lowered photosynthetic activity, and thus ATP synthesis rates, the gravitational force exceeds the net upward propulsion, causing the cells at the top interface to sediment.
In turn, gyrotaxis forces the cells to accumulate in highly localized downwelling flows, finally leading to the emergence of spatially regular and temporally stationary bioconvective plumes.

In summary, confined photosynthetically active microbial suspensions display reversibly light-switchable coherent flows. For unfavourable natural conditions in which - at least locally and temporally - both oxygen and light may be deprived, this mechanism might be advantageous for the survival and propagation of the entire microbial community.
In technological settings involving, e.g., the synthesis of molecular hydrogen, the findings may guide new design principles for photobioreactors in the fields of renewable energies and molecular farming.

\section*{Materials and Methods}
\subsection*{Cell cultivation}
Wild-type \textit{C.~reinhardtii} cells, strain SAG~11-32b provided by the G\"ottingen Algae Culture Collection (SAG, Göttingen, Germany), were grown axenically in Tris-Acetate-Phosphate (TAP, Gibco\texttrademark) medium on 12\,h$-$12\,h day-night cycles at temperatures of $24\,^\circ$C and $22\,^\circ$C, respectively, in a Memmert IPP260ecoplus incubator.
Cultures of $100$\,mL were prepared in $250$\,mL Erlenmeyer flasks and placed on an orbital shaker (Advanced Mini Shaker 15, VWR) with an orbit diameter of $15$\,mm at $100$\,rpm.
The daytime light intensity in the incubator was $(1-2)\cdot 10^{19}$\, photons m$^{-2}$ s$^{-1}$ using cool white (6500~K) LEDs, and zero during the nighttime. 
Vegetative cells were taken from the cultures in the logarithmic growth phase during the daytime on the third or fourth day after starting the incubation.
In order to increase the cell density, the culture was centrifuged for $10$\,min at an acceleration of $100$\,g, the excess fluid was removed, and the pellet of cells was resuspended in fresh TAP medium. 
Since cells may deflagellate due to the mechanical stress during the centrifugation, the cell suspension was allowed to rest for $1.5$\,hours to allow for regrowing their flagella \cite{harris2023}.
A hemocytometer (Neubauer-improved with double net ruling) was used to manually measure the cell density, and then dilute the suspension to the final desired density of $(8\pm1)\cdot10^7$\,cells/mL.

\subsection*{Bioconvection setup}
All bioconvection experiments were performed with a custom-made setup.
The imaging of the samples was either performed in horizontal or vertical orientation depending on the type of compartment used, see Fig.~S1 in the SI. 
The experiments were performed under red light to exclude both phototaxis \cite{berthold2008channelrhodopsin} and adhesion to surfaces \cite{kreis2018adhesion,catalan2023light}. 
An LED light source with center wavelength of 660\,nm and a full width at which the intensity is half of the maximum (FWHM) of 20\,nm was used to illuminate the sample (M660L4, Thorlabs).
An optical diffuser was used to homogenize the light from the light source (DG20-600, Thorlabs), while neutral density filters with optical densities of 1 or 2 were used to expand the range of light intensities that can be used in the experiment (NE2R10B and NE2R20B, Thorlabs).
A condenser lens was used to ensure K\"{o}hler illumination, which provides reproducible and homogeneous illumination conditions (LA1740, Thorlabs). 
The image is focused using a macro zoom lens ($12.5-75$\,mm f$1.8$, Avenir, Japan) and recorded at $3$\,fps with a CMOS camera (Grasshopper3, GS3-U3-51S5M, Flir) with an 8-bit depth.
The camera sensor has a size of 2448x2048 pixels with a pixel size of $3.45$\,\textmu m.
For both experimental configurations (bioconvection and cell motility setup) the light intensity was calibrated using a Thorlabs PM100D powermeter (Thorlabs S130C photodiode power sensor) for monochromatic light.
%The light intensity is found to decrease at the edge of the compartment by up to $15\%$. 
%The values reported in the paper are the ones measured at the center of the compartment.

The compartments for containing the living suspensions were made of custom-made stainless steel parts in between two soda lime glass slides (Marienfeld Superior), as shown in Fig.~S1(b,c) in the SI.
In order to ensure that the compartments are airtight, the parts were manually clamped together.
Functionalizing the glass slides with a thin polydimethylsiloxane (PDMS) coating has shown to improve the  sealing with the steel component.
%However, the results with the PDMS film are indistinguishable with the plain glass slides.
%As a result, we have used both types of surfaces.
For this coating, Sylgard 184 PDMS (Dow Chemical) was mixed with a 10:1 by weight ratio of base to curing agent. 
The mixture of PDMS was spin-coated using a commercial spin-coater (WS-650MZ-23NPPB, Laurell Technologies Corp., US) onto a glass slide at $950$\,rpm for $5$\,minutes, to achieve the final thickness of about $21\pm1$\,\textmu m. 
After spin-coating, the glass slide was immediately placed on a hot plate for $4$ minutes at about $95\,^\circ$C to accelerate the cross-linking process. 
Afterwards, the glass slides were placed in an oven at $75^\circ$C for 2\,hours to complete the polymerization process.
Since the PDMS coating renders the glass surface hydrophobic, it impedes the filling of the compartments with the suspension without air bubbles.
For this reason, the coated glass slides are plasma cleaned with atmospheric air for $30$\,s (Atto plasma cleaner, Diener, Germany), which renders the surface hydrophilic.
All data provided in Fig.~\ref{fig:fig3} in quasi-2D compartments were obtained for light intensities $I\ge 1.7\cdot10^{17}$\, photons m$^{-2}$ s$^{-1}$.
For lower light intensities, bioconvective plumes in the quasi-2D compartments did emerge but not achieve a clear steady state.

The experiments for bioconvection in aerobic conditions, shown in Fig.\ref{fig:fig1}a, were performed using conventional polystyrene petri dishes with a $55$\,mm inner diameter (VWR). 
The petri dishes were plasma cleaned with atmospheric air for $30$\,s to render the surface hydrophilic, after which the cell suspension was carefully pipetted in. 
Finally, the suspension was illuminated with a halogen cold light source (LK 1500 LCD, Schott).

\subsection*{Cell motility setup}
The cell motility experiments were performed using an Olympus IX81 inverted microscope.
The condenser was always adjusted to ensure K\"{o}hler illumination.
An interference bandpass filter with center wavelength of 671\,nm and a FWHM of 10\,nm ensured consistent illumination of the sample with red light in these experiments.
A 4x objective was used in conjunction with the magnification changer set at 1.6x resulting to a total of 6.4x magnification. 
Videos were recorded at $30$\,fps with a CMOS camera (Grasshopper3, GS3-U3-41C6M, Flir).

The airtight compartments are the same as the 3D compartments, shown in Fig.~S1b in the SI, but with a height of $25$\,\textmu m and a $3$\,mm diameter.
The air-permeable compartments were made using the spin-coated PDMS of a height of $21\pm 1$\,\textmu m as described in the previous sections.
A $3$\,mm in diameter circular punch (Harris Uni-Core) was used to produce the desired circular compartment.
In addition, the top glass slide was exchanged for a $1$\,mm thick solid PDMS to allow for sufficient air exchange.

The circular Hough transform was used to detect the cells \cite{yuen1990comparative}, which requires only partial detection of the cell's edge to identify it. The circular Hough transform requires a previous knowledge of the radius of the cells. In our case, we used a range of radii between $2.6$\,\textmu m and $5.5$\,\textmu m \cite{catalan2023light}. 
The trajectories of the detected cells are determined using a Matlab-based code by Blair \& Dufresne \cite{blair2008matlab}, which requires the maximum distance that cells can be displaced between consecutive frames. 
We chose this parameter to be $7$\,\textmu m, which results in a maximum instantaneous velocity of $233$\,\textmu m\,s$^{-1}$. 
Finally, the cell velocity is calculated as the displacement of the tracked cells over one frame.

\subsection*{Fourier analysis \& particle imaging velocimetry}
Since the wavelength of the instability rarely matches exactly the discretization of the gathered data, spectral leakage occurs, where the peak of the power spectrum leaks into nearby wavelengths \cite{lyon2009discrete}.
To limit this effect, we scale the intensity in our images, $J(x,z,t)$, with the Hamming window, $H(x,z)$ \cite{jackson1991selection}.
The resultant power spectrum is given as:
\begin{equation}\label{eq:FT}
S(q_x,q_y,t) = \frac{|\mathcal{F}[J(x,y,t)\cdot H(x,y)]|}{|\mathcal{F}[H(x,y)]|},
\end{equation}
where $\mathcal{F}$ is the Fourier transform operator, while $q_x$ and $q_y$ are the wavenumbers along the horizontal and vertical directions of the images.
In the case of the 3D compartments, $S(q_x,q_y,t)$ exhibits azimuthal symmetry, see Fig.~S3a in the SI for a representative example.
Consequently, we use the azimuthal average for our analysis, i.e.\ $S(q,t) = \langle S(q_x,q_y,t)\rangle_\theta$, where $\theta$ is the azimuthal angle.
An example of $S(q,t)$ in a 3D compartment is provided in Fig.~S3b in the SI.
Due to the limited height in quasi-2D compartments, we use instead a 1D signal for the Fourier transform along the length of the compartment. 
The images, $J(x,z)$, were first converted to a 1D signal as $J(x) = \langle J(x,z)\rangle_z$.
Using this 1D signal the same process as in eq.~\ref{eq:FT} was performed using a 1D Hamming window.

For the PIV analysis, the Matlab-based and open access software PIVlab was used \cite{thielicke2021particle}.
The images are pre-processed using an enhanced local contrast algorithm (CLAHE) with a window size of $1.5$\,mm \cite{reza2004realization}.
The images were analyzed using the fast Fourier transform window deformation algorithm with progressive refinement of the interrogation window size ($0.24,0.19,0.15$\,mm) and step size ($0.12,0.1,0.7$\,mm).
Finally, we average the cell flow field over all the processed frames of the stable plume.

\subsection*{Cell density estimation}
To quantify the cell density we use the Beer-Lambert law that estimates the concentration of light-absorbing matter by measuring the attenuation of light \cite{blanken2016predicting}.
Specifically, the Beer-Lambert law is given by
\begin{equation}\label{eq:BL1}
    \rho = -\alpha \log\left(\frac{I_\mathrm{T}}{I_0}\right),
\end{equation}
where $\rho$ is the local cell density, and $I_\mathrm{T}$ and $I_0$ the transmitted and incident light intensity, respectively.
The parameter $\alpha$ depends solely on the path length of the light and the attenuation coefficient of the cells.
Since all quasi-2D experiments exhibit equivalent width, thus the same path length, $\alpha$ can be considered a constant for these experiments.
During experiments, we measure the transmitted light intensity $I_\mathrm{T}$, and thus two parameters in eq.~\ref{eq:BL1} remain unknown, namely $\alpha$ and $I_0$.
Since we perform all experiments with the same cell density, we can use this reduce the unknown parameters.
Given eq.~\ref{eq:BL1}, we can calculate the average cell density as
\begin{equation}\label{eq:BL2}
    \rho_0=\frac{1}{A}\iint_A\rho\,dx\,dz, = -\alpha\left\langle\log\left(\frac{I_\mathrm{T}}{I_0}\right)\right\rangle
\end{equation}
where $A$ is the area over which we view the suspension.
Using eqs.~\ref{eq:BL1} and \ref{eq:BL2} and that $I_0$ is considered a constant within the processed area, we estimate the relative cell density in the sample as
\begin{equation}\label{eq:LB3}
    \left(\frac{\rho-\rho_0}{\alpha}\right) = \left\langle\log\left(I_\mathrm{T}\right)\right\rangle - \log\left(I_\mathrm{T}\right).
\end{equation}
The Beer-Lambert law is based on the absorption of light, and it does not take into account light scattering.
Even though it provides a good approximation of the cell density \cite{blanken2016predicting}, it deviates at high cell densities, where scattering becomes prominent.
This transition is shifted to higher cell densities in the case of photosynthetic organisms specifically due to the high absorbance capacity of photosynthetic pigments at specific wavelengths as is in our case \cite{myers2013improving}.
Nevertheless, we note that using this method might result in an underestimation of the cell density in dense regions.

\subsection*{Chlorophyll auto-fluorescence}
Chlorophyll a/b exhibit two major peaks of absorption around $470$\,nm and $670$\,nm, while they emit light at higher wavelengths in the range of $650-750$\,nm via auto-fluorescence \cite{lamb2018chlorophyll}.
Auto-fluorescence measurements were performed along the line of the cell motility experiments using a spin-coated PDMS compartment with a glass slide cover. 
In these experiments, a $671$\,nm bandpass filter with a $10$\,nm FWHM was placed between the light source and the sample, while a $700$\,nm long-pass filter right after the sample allows only the fluorescent light to be detected by the camera.
For consistency, the camera settings were kept constant for all experiments. 
Two experimental measurements were performed for each light intensity: one with the sample and one without.
Each measurement consisted of a series of $300$ images captured at $33$\,fps using a sCMOS camera (Iris 9, Photometrics) with relatively high quantum efficiency ($\sim66$\% in red light) and 16-bit depth.
The camera sensor has a size of 2960x2960 pixels with a pixel size of $4.25$\,\textmu m.
Images acquired without any sample were used to subtract the background noise in the actual measurements. 
Finally, the average auto-fluorescence signal was then calculated for each light intensity, which is displayed in Fig.~\ref{fig:fig4}c.

\section*{Acknowledgments}

\begin{acknowledgments}
The authors sincerely thank the Göttingen Algae Culture Collection (SAG) for providing the \textit{Chlamydomonas reinhardtii} wild-type strain SAG 11-32b and Peter Hegemann (Humboldt University of Berlin) for providing the channelrhodopsin-1,2 double knockout mutant (CC-5679) in this background. We thank M.\ Lorenz, S. Herminghaus, M.\ Wilczek, C.-M.\ Koch and H.\ Stark for stimulating scientific discussions and K.\ Oetter for technical assistance. Funding is acknowledged from the Max Planck Society.
\end{acknowledgments}

\end{document}